\documentclass[aps,prd,preprint,showpacs,superscriptaddress]{revtex4}

\begin{document}
\title{Dark matter from the scalar sector of 3-3-1 models without exotic electric charges}
\author{Simonetta Filippi}
\affiliation{I.C.R.A.-International Center for Relativistic Astrophysics- at Physics Department, University of Rome ``La  Sapienza'', P.le A. Moro 5, 00185 Rome, Italy}
\affiliation{University CBM, Via E. Longoni 83, 00155 Rome, Italy}
\author{William A. Ponce} 
\affiliation{Instituto de F\'\i sica, Universidad de Antioquia,
A.A. 1226, Medell\'\i n, Colombia}
\author{Luis A. S\'anchez}
\affiliation{I.C.R.A.-International Center for Relativistic Astrophysics- at Physics Department, University of Rome ``La  Sapienza'', P.le A. Moro 5, 00185 Rome, Italy}
\affiliation{Escuela de F\'\i sica, Universidad Nacional de Colombia,
A.A. 3840, Medell\'\i n, Colombia}


\begin{abstract}
It is shown that three $SU(2)$ singlet neutral scalars (two CP-even and one CP-odd) in the spectrum of models based on the gauge symmetry $SU(3)_c\otimes SU(3)_L\otimes U(1)_X$, which do not contain exotic electric charges, are realistic candidates for thermally generated self-interacting dark matter in the Universe, a type of dark matter that has been recently proposed in order to overcome some difficulties of collisionless cold dark matter models at the galactic scale. These candidates arise without introducing a new mass scale in the model and/or without the need for a discrete symmetry to stabilize them, but at the expense of tuning several combinations of parameters of the scalar potential. \end{abstract}


\maketitle

The inflationary model plus the cold dark matter model supplemented by a cosmological constant have been very successful in explaining the observed structure of the Universe on large scales. Notwithstanding, in the past years very high resolution N-body simulations have shown that collisionless cold dark matter (CDM) models produce far too much substructures on galactic scales as compared with observations \cite{nbody1}, and predict dark matter halo density profiles that diverge at the centre in disagreement with the constant density cores observed in late-type dwarf and low surface brightness galaxies \cite{nbody2}.

A possible way out to these problems is to assume that the dark matter particles are self-interacting with a large scattering cross-section but negligible annhilation or dissipation \cite{Spergel}. However, self-interacting CDM models lead to spherical halo centers in clusters which is not in agreement with ellipsoidal centers indicated by strong gravitational lensing data \cite{Yoshida} and have difficulties to reconcile the number of satellites in the Galactic halo with the observed number of dwarf galaxies in the Local Group \cite{astroph}. In spite of these difficulties, self-interacting dark matter models are well-motivated as alternative models.

The unsatisfactory \textit{ad hoc} addition of just one singlet neutral scalar field, and a discrete symmetry in order to guarantee its stability, as the simplest possible renormalizable extension of the Standard Model that includes non-baryonic self-interacting dark matter \cite{mcdonald, burgess}, has motivated several authors to investigate the possibility for such a class of dark matter arising naturally from realistic gauge models. In particular, Refs.~\cite{FreTo} and \cite{LoLa} have studied this issue in the context of the Standard Model (SM) extension based on the $SU(3)_c \otimes SU(3)_L \otimes U(1)_X$ gauge symmetry (the 3-3-1 model). In Ref.~\cite{FreTo}, by introducing a new very small energy scale ($\sim 10^{-7}$ GeV), a physical CP-odd neutral field satisfaying the Spergel-Steinhardt bound \cite{Spergel} for self-interacting dark matter has been identified in the scalar sector of the Pisano-Pleitez-Frampton model (PPF) \cite{pf}, which is a model that includes quarks with exotic electric charges $-4/3$ and $5/3$ and double charged Higgs and gauge bososns. In Ref.~\cite{LoLa} and after imposing by hand a discrete symmetry, two scalar fields, one CP-odd and one CP-even, have been identified as realistic candidates for self-interacting dark matter in the version of the model referred to as ``3-3-1 model with right-handed neutrinos'' \cite{foot}, which is a model where the particles have only ordinary electric charges.

Our goal in this letter is to search for realistic candidates for thermally generated self-interacting dark matter arising from the scalar sector of 3-3-1 models without exotic electric charges, without the need of imposing an extra symmetry and/or without introducing a new mass scale, and taking care of the consistency of the full scalar spectrum of the theory. Thermally generated singlet scalars with MeV range masses and $O(M_{Pl}^{-1})$ couplings to matter was proposed for the first time as self-interacting dark matter in~\cite{mcdonald}, and has been recently proposed as a candidate for explaining the observed fluxes of 511 KeV photons from the Galactic bulge \cite{pospelov}. Clearly, this range of masses is not in the 1 GeV - 1 TeV range preferred by most models of weakly interacting massive particles (WIMPS) \cite{wandelt}. 

The 3-3-1 extension of the SM has received special attention over the last decade since several models can be constructed so that when anomaly cancellation takes place between the fermion families (three-family models), and not family by family as in the SM (one-family models), these models predicts that the number $N_f$ of families must be an integer multiple of the number $N_c$ of colors of $SU(3)_c$. Moreover, since $SU(3)_c$ asymptotic freedom requires $N_c<5$, it follows that $N_f=N_c=3$. In this way these models provide a possible answer to the question of the number of fermion families in nature \cite{pf,valle}.

All possible 3-3-1 models without exotic electric charges are presented in Ref.~\cite{pfs}. These models share in common not only the same gauge boson sector but also the same scalar sector. 
For the 3-3-1 models both, $SU(3)_L$ and $U(1)_X$ are anomalous, so special combination of multiplets must be used in each particular model in order to cancel the several possible anomalies, and end with physical acceptable models. It is shown in Ref.~\cite{pfs} that by demanding cancellation of the triangle anomalies: $[SU(3)_L]^3$, $[SU(3)_c]^2U(1)_X$, $[SU(3)_L]^2U(1)_X$, $[grav]^2U(1)_X$ (the gravitational anomaly), and $[U(1)_X]^3$, only ten different models without exotic electric charges can be constructed. Two of them are one-family structures and have been studied in Refs.~\cite{spm} and ~\cite{mps}. The other eight models are three-family models. One of them is the already mentioned ``3-3-1 model with right-handed neutrinos'' \cite{foot}, another one has been analyzed in~\cite{ozer}, two more have been studied in~\cite{sher}, and the remainig four models have not yet been studied in the literature, as far as we know.

Now, if our aim is to break the symmetry in the way
\begin{eqnarray} \nonumber
SU(3)_c\otimes SU(3)_L\otimes U(1)_X & & \nonumber \\
\longrightarrow& SU(3)_c\otimes SU(2)_L\otimes U(1)_Y& \nonumber \\
\longrightarrow& SU(3)_c\otimes U(1)_Q,& \nonumber \\ \label{break}
\end{eqnarray}
and at the same time give masses to the fermion fields in
the model, the following three Higgs scalars must be introduced \cite{foot}: $\Phi_1(1,3^*,-1/3)=(\phi_1^-,\phi_1^{0},\phi_1^{\prime 0})$ with Vacuum Expectation Value (VEV) $\langle\Phi_1\rangle^T=(0,0,V)$;
$\Phi_2(1,3^*,-1/3)=(\phi_2^-,\phi_2^{0},\phi_2^{\prime 0})$ with VEV
$\langle\Phi_2\rangle^T=(0,v,0)$, and
$\Phi_3(1,3^*,2/3)=(\phi_3^0,\phi_3^+,\phi_3^{'+})$ with VEV
$\langle\Phi_3\rangle^T=(u,0,0)$, with the hierarchy
$V \gg v\sim u\sim 174$ GeV, the electroweak breaking scale. The consistency with the low energy phenomenology requires $\sqrt{u^2+v^2}=174$ GeV \cite{foot}, and by using experimental results from the CERN LEP, SLAC Linear Collider and atomic parity violation data, a lower bound on the mass scale $V$ have been calculated in Refs.~\cite{spm, mps}; generically, $V \geq 1$ TeV.

The most general renormalizable scalar potential which includes $\Phi_1$, $\Phi_2$ and $\Phi_3$ can be written in the following form
\begin{widetext}
\begin{eqnarray}\nonumber
V(\Phi_1,\Phi_2,\Phi_3)&=&
\mu^2_1\Phi_1^\dagger\Phi_1 + \mu^2_2\Phi_2^\dagger\Phi_2 + \mu^2_3\Phi_3^\dagger\Phi_3 + {\mu^2_4\over 2}(\Phi_1^\dagger\Phi_2 + h.c.)
\\ \nonumber
& & +\lambda_1(\Phi_1^\dagger\Phi_1)^2 +\lambda_2(\Phi_2^\dagger\Phi_2)^2 +\lambda_3(\Phi_3^\dagger\Phi_3)^2 +{\lambda_4\over 2} [(\Phi_1^\dagger\Phi_2)(\Phi_1^\dagger\Phi_2)+h.c.]\\ \nonumber
& & +\lambda_5(\Phi_1^\dagger\Phi_1)(\Phi_2^\dagger\Phi_2) + \lambda_6(\Phi_1^\dagger\Phi_1)(\Phi_3^\dagger\Phi_3) +
\lambda_7(\Phi_2^\dagger\Phi_2)(\Phi_3^\dagger\Phi_3) \\ \nonumber
& & +\lambda_8(\Phi_1^\dagger\Phi_2)(\Phi_2^\dagger\Phi_1) +\lambda_9(\Phi_1^\dagger\Phi_3)(\Phi_3^\dagger\Phi_1) +\lambda_{10}(\Phi_2^\dagger\Phi_3)(\Phi_3^\dagger\Phi_2) \\ \nonumber
& & + \frac{1}{2}[\lambda_{11}(\Phi^\dagger_1\Phi_1)(\Phi^\dagger_1\Phi_2)+ \lambda_{12}(\Phi^\dagger_2\Phi_2)(\Phi^\dagger_1\Phi_2)+\lambda_{13}(\Phi^\dagger_3\Phi_3)(\Phi^\dagger_1\Phi_2) \\ \label{vphi} 
& &+\lambda_{14}(\Phi^\dagger_1\Phi_3)(\Phi^\dagger_3\Phi_2)
-f\epsilon_{ijk}\Phi^i_1\Phi^j_2\Phi^k_3 + h.c.], 
\end{eqnarray}
\end{widetext}
which is the same scalar potential studied in Refs.~\cite{long,dmo}. In this equation the $\mu$'s and $f$ are coupling constants with dimension of mass and the $\lambda$'s are dimensionless.

For the sake of simplicity we assume that the VEV are real and, for convenience in reading, we rewrite the neutral scalar fields as

\begin{eqnarray}\nonumber 
\phi_1^0&=&\frac{\phi^0_{1R}+i\phi^0_{1I}}{\sqrt{2}},\\ \nonumber
\phi_2^{\prime 0}&=&\frac{\phi^{\prime 0}_{2R} 
+i\phi^{\prime 0}_{2I}}{\sqrt{2}},\\ \nonumber
\phi_1^{\prime 0}&=&V+\frac{\phi^{\prime 0}_{1R}+i\phi^{\prime 0}_{1I}}{\sqrt{2}},\\ \nonumber
\phi_2^{0}&=&v+\frac{\phi^0_{2R}+i\phi^0_{2I}}{\sqrt{2}}, \\
\label{neutras}
\phi_3^0&=&u+\frac{\phi^0_{3R}+i\phi^0_{3I}}{\sqrt{2}}. 
\end{eqnarray}

In the literature, a real part $\phi_{R}$ is called a CP-even scalar, and an imaginary part $\phi_{I}$ is called a CP-odd scalar or pseudoscalar field.
 
A complete study of the scalar mass spectrum for the type of 3-3-1 models we are interested in has been done in Ref.~\cite{dmo}, in the limit $f \simeq V \gg u,v$. In this limit, the couplings of one of the 3-3-1 CP-even scalars to the SM fermions and gauge bosons become identical to the couplings of the SM Higgs, and the mass spectrum for the scalar fields both in the charged and in the neutral (CP-even and CP-odd) sectors, is as follows \cite{dmo}.
 
For the charged scalar sector the square mass matrix has two eigenvalues equal to zero equivalent to four would be Goldstone bosons $(G_1^\pm ,G_2^\pm)$. Now, with $C_\alpha \equiv \cos \alpha =u/\sqrt{u^2+v^2}$, $S_\alpha\equiv \sin \alpha =v/\sqrt{u^2+v^2}$, and in the limit $f \simeq V \gg u,v$, the physical charged scalars are
\begin{eqnarray}\nonumber
H^{\pm}_1 &\simeq& \phi^{\prime \pm}_3, \nonumber \\
H^{\pm}_2 &\simeq& C_\alpha \phi^\pm_2 + S_\alpha \phi^\pm_3, \label{char}
\end{eqnarray}
with squared masses at the TeV scale given by
\begin{eqnarray}\nonumber
m^2_{H^{\pm}_1} &\simeq& \lambda_9 V^2 -\frac{1}{2}fV\frac{v}{u}, \nonumber \\ 
m^2_{H^{\pm}_2} &\simeq& \frac{1}{2}fV\left(\frac{u}{v}+\frac{v}{u}\right),
\label{physch}
\end{eqnarray} 
respectively.

For the CP-even scalars the square mass matrix has one eigenvalue equal to zero corresponding to another would be Goldstone boson $G_3^0 \simeq -\phi^0_{1R}$. Defining
\begin{eqnarray}\nonumber
\cos 2\beta &=& \frac{2(\lambda_8+\lambda_4-4\lambda_1)v+u}{\sqrt{[2(\lambda_8+\lambda_4-4\lambda_1)v+u]^2+16\lambda^2_{11}v^2}}, \\ \nonumber
\sin 2\beta &=& \frac{4\lambda_{11}v}{\sqrt{[2(\lambda_8+\lambda_4-4\lambda_1)v+u]^2+16\lambda^2_{11}v^2}},
\end{eqnarray}
and in the limit $f \simeq V \gg u,v$, the physical CP-even scalars and their squared masses are
\begin{eqnarray}\nonumber
H^0_1 &\simeq& C_\alpha \phi^0_{2R}-S_\alpha \phi^0_{3R}, \qquad m^2_{H^0_1} \simeq \frac{1}{2}fV; \nonumber \\
H^0_2 &\simeq& S_\alpha \phi^0_{2R}+C_\alpha \phi^0_{3R}, \qquad m^2_{H^0_2} \simeq uv; \nonumber \\
H^0_3 &\simeq& C_\beta \phi^{\prime 0}_{1R}-S_\beta \phi^{\prime 0}_{2R}, \qquad m^2_{H^0_3} \simeq \frac{1}{4}\Lambda_1 V^2; \nonumber \\
H^0_4 &\simeq& S_\beta \phi^{\prime 0}_{1R}+C_\beta \phi^{\prime 0}_{2R}, \qquad m^2_{H^0_4} \simeq \frac{1}{4}\Lambda_2 V^2, \label{cpeve}
\end{eqnarray}
where 
\begin{eqnarray}\nonumber
\Lambda_{1,2}&=&\Lambda_+ \pm \sqrt{\Lambda^2_- +4\lambda^2_{11}}>0, \nonumber \\
\Lambda_\pm &=& (fu/2vV)+\lambda_4 +\lambda_8 \pm 4\lambda_1, \label{coef}
\end{eqnarray}
and  $C_\beta$, $S_\beta$ stand for $\cos \beta$, $\sin \beta$, respectively.

Notice that $H^0_3$ and $H^0_4$ are mainly mixtures of $SU(2)$ singlets and that their masses depend on the combinations of parameters $\Lambda_1$ and $\Lambda_2$ which are not fixed by the experiment, except for $\sqrt{u^2+v^2}=174$ GeV. So, $m_{H^0_3}$ and $m_{H^0_4}$ can in principle be tuned in order to have two light CP-even scalars in the low energy spectrum of the model. Notice also that $H^0_1$ and $H^0_2$ are mainly mixtures of members of $SU(2)$ doublets. Besides, it has been shown in Ref.~\cite{dmo} that $H^0_2$ is to be identified with the SM Higgs [See Eq.~(\ref{tri}) below].

For the CP-odd scalars the square mass matrix has three zero eigenvalues associated with the three additional would-be Goldstone bosons: $G^0_4 \simeq \phi^0_{1I}$,
$G^0_5 \simeq \phi^{\prime 0}_{1I}$, and 
$G^0_6 \simeq -S_\alpha \phi^{0}_{2I}+ C_\alpha \phi^0_{3I}$. The nonzero eigenvalues correspond to the two physical states
\begin{eqnarray}\nonumber
h^0_1 &\simeq& C_\alpha \phi^{0}_{2I}+ S_\alpha \phi^0_{3I}, \nonumber \\
h^0_2 &\simeq& \phi^{\prime 0}_{2I}; \label{sodd}
\end{eqnarray}
with squared masses given by
\begin{eqnarray}\nonumber
m^2_{h^0_1}&\simeq& \frac{1}{4}f\frac{u^2+v^2}{uv}V, \nonumber \\
m^2_{h^0_2}&\simeq& \frac{1}{4}\frac{[fuV - 2v(\lambda_4 - \lambda_8)V^2]}{v} \nonumber \\
 &\simeq& \frac{1}{4}\frac{[u - 2v(\lambda_4 - \lambda_8)]V^2}{v}, \label{cpodd}
\end{eqnarray}
respectively.

Also in this sector, the scalar $h^0_2$ is mainly a SM singlet and its mass depends on the difference $u - 2v(\lambda_4 - \lambda_8)$, where $\sqrt{u^2+v^2}=174$ GeV. Since the value of $\lambda_4 - \lambda_8$ is not fixed by the experiment, we can tune the value of $m_{h^0_2}$ so that we also have a very light CP-odd scalar in the low energy spectrum of the model. Notice that $h^0_1$ is heavy and is mainly a mixture of electrically neutral components of $SU(2)$ doublets.

It is worth to say that, as required by the consistency of the model, the original eighteen degrees of freedom in the scalar sector have become eight would be Goldstone bosons (four electrically neutral and four charged), and ten physical scalar fields, six neutrals and four charged ones.

Since the CP-even scalars $H^0_3$ and $H^0_4$ and the CP-odd scalar $h^0_2$ are mainly SM singlets, they couple very weakly to the scalars $h^1_0$, $H^1_0$ and $H^0_2$ (which are mixtures of electrically neutral components of $SU(2)$ doublets) and should couple very weakly to the SM fermions and gauge bosons. In fact, the trilinear couplings between the physical scalars of the model and the SM gauge bosons have been calculated in Ref.~\cite{dmo} and the corresponding Lagrangian, in the limit $f \simeq V \gg u,v$, is
\begin{eqnarray}\nonumber
L^{SM}_{trilinear}&=& \left(gM_{W^\pm}W^{\mu -}W^+_\mu +\frac{g}{2C_W}M_Z Z^\mu Z_\mu \right) H^0_2 \\ \nonumber
 & & -[(eM_{W^\pm}A^\mu + gM_Z S^2_W Z^\mu) W^+_\mu G^-_1 + h.c.] \\ \nonumber
 & & -\frac{g}{2}\Bigg\{(p-k)^\mu W^+_\mu H^-_2 H^0_1 \\ \label{tri}
 & & +i(p-r)^\mu W^+_\mu H^-_2 h^0_1 + h.c.\Bigg\},
\end{eqnarray}
where the scalars $h^0_2$, $H^0_3$ and $H^0_4$ are absent and from which it is clear that the couplings of $H^0_2$ coincide with the ones of the SM Higgs.

These features of $h^0_2$, $H^0_3$ and $H^0_4$ make them stable without the need of imposing an extra symmetry, and make also them suitable candidates for self-interacting dark matter in the Universe. Let us see.

A realistic candidate for self-interacting dark matter must have mean free path $\Lambda$, at the solar radius in a typical galaxy, in the range $1$ kpc $\leq \Lambda \leq 1$ Mpc, where $\Lambda = 1/(n \sigma)$. $n =\rho/m_{h^0}$ is the number density of a generic singlet scalar field $h^0$, $\sigma$ is the cross section for the self-interaction  $h^0h^0 \rightarrow h^0h^0$, and $\rho$ is the mean density of a typical galaxy ($\rho = 0.4$ GeV/cm$^3$ \cite{Spergel}). The bound on $\Lambda$ can be translated into a bound on the quotient $\sigma/m_{h^0}$, namely 
\begin{equation}\label{bound}
2\; \mathrm{x}\; 10^3\; \mathrm{GeV}^{-3} \leq \frac{\sigma}{m_{h^0}} \leq 3\;  \mathrm{x}\; 10^4\; \mathrm{GeV}^{-3}.
\end{equation}
Near the threshold ($s \simeq 4 m^2_{h^0}$) the cross section $\sigma$ is roughly given by
$\sigma = \lambda^2_S/(16 \pi s) \simeq \lambda^2_S/(64 \pi m^2_{h^0})$,
where $\lambda_S$ is the coupling constant for self-interaction. 
The value of $\lambda_S$ for $h^0_2$, $H^0_3$ and $H^0_4$, obtained from the scalar potential written in terms of physical fields, are given respectively by 
\begin{eqnarray}\nonumber
\lambda_{S1} &=& \lambda_2, \nonumber \\
\lambda_{S2} &=& \lambda_1\cos^4\beta+\lambda_2\sin^4\beta \nonumber \\
 & & +\frac{1}{4}(\lambda_4+\lambda_5+\lambda_8)sin^22\beta \nonumber \\
 & & -\frac{1}{2}(\lambda_{11}\cos^2\beta+\lambda_{12}\sin^2\beta)
\sin2\beta, \nonumber \\
\lambda_{S3} &=& \lambda_1\sin^4\beta+\lambda_2\cos^4\beta \nonumber \\
 & & +\frac{1}{4}(\lambda_4+\lambda_5+\lambda_8)sin^22\beta \nonumber \\
 & & +\frac{1}{2}(\lambda_{11}\sin^2\beta+\lambda_{12}\cos^2\beta)
sin2\beta. \label{lambdas}
\end{eqnarray}
Let us write the mass $m_{h^0}$ of each one of our dark matter candidates in the form
\begin{equation}\label{msings}
m_i=\frac{1}{2}\sqrt{\frac{\Delta_i}{v}}V,
\end{equation}
where $i=1,2,3$, $\Delta_1 = \vert u - 2v(\lambda_4 - \lambda_8)\vert$, 
$\Delta_2 = v \vert \Lambda_1\vert$, and $\Delta_3 = v \vert\Lambda_2\vert$, for $h^0_2$, $H^0_3$ and $H^0_4$, respectively, with $\Lambda_{1,2}$ as defined in Eq.~(\ref{coef}).

With $\lambda_{Si}=0.5$ (a natural value as compared to the SM Higgs self-coupling), $V= 1$ TeV, $v = 123$ GeV, and if we tune each one of the $\Delta_i$ so as to lie in the range $5.90 \times 10^{-9}$ GeV 
$\leq \Delta_i \leq 3.57 \times 10^{-8}$ GeV, we obtain the required Spergel-Steinhardt bound in Eq.~(\ref{bound}) and, for each one of the dark matter candidates provided by the present model, a tree-level mass in the range: $3.46$ MeV $\leq m_i\leq 8.52$ MeV, a typical value for thermally generated self-interacting dark matter scalars \cite{mcdonald}. The range of values for each $\Delta_i$ can be tuned with a variety of values of the parameters involved in their definitions.
 
The scalars $h^0_2$, $H^0_3$ and $H^0_4$ with a mass in the range of a few MeV, are non-relativistic in the decoupling era ($T_{dec}\sim 1$ eV), then, for a SM Higgs boson with a mass of the order of $100$ GeV the dominant contribution to the cosmic density of these scalars comes from the decay of thermal equilibrium scalar $H^0_2$ (the scalar of the model to be identified with the SM Higgs) to $h^0h^0$ pairs at temperatures less than the temperature $T_{EW}$ of the electroweak phase transition ($T_{EW}\geq 1.5m_{H^0_2}$) \cite{mcdonald}. The trilinear couplings of $h^0_2$, $H^0_3$ and $H^0_4$ to $H^0_2$, also obtained from the scalar potential in Eq.~(\ref{vphi}) written in terms of physical fields, have the following strengths:
\begin{eqnarray}\nonumber
H^0_2h^0_2h^0_2: 
& \Gamma_1=\frac{1}{\sqrt{2}}\Big\{v\lambda_2 \sin \alpha +\frac{u}{2}\lambda_7 \cos \alpha\Big\}, \nonumber \\
H^0_2H^0_3H^0_3: 
& \Gamma_2=\frac{v}{4\sqrt{2}}\sin\alpha\Big\{4\lambda_2 
\sin^2 \beta +2\lambda_5 \cos^2 \beta \nonumber \\
 & -\lambda_{12}\sin2\beta\Big\} \nonumber \\
 & +\frac{u}{4\sqrt{2}}\cos\alpha\Big\{2\lambda_6\cos^2\beta+2\lambda_7
\sin^2\beta \nonumber \\
 & -\lambda_{13}\sin2\beta\Big\}, \nonumber \\
H^0_2H^0_4H^0_4: 
 & \Gamma_3=\frac{v}{4\sqrt{2}}\sin\alpha
\Big\{4\lambda_2 \cos^2 \beta +2\lambda_5 \sin^2 \beta \nonumber \\
 & +\lambda_{12}\sin2\beta\Big\} \nonumber \\
 & +\frac{u}{4\sqrt{2}}\cos\alpha\Big\{2\lambda_6\sin^2\beta+2\lambda_7
\cos^2\beta \nonumber \\
 & +\lambda_{13}\sin2\beta\Big\}. \label{acoples}
\end{eqnarray}
The cosmic density of light gauge singlet scalars from the $H^0_2$ decay has been calculated in Ref.~\cite{mcdonald} and is given by
\begin{equation}\label{cosmic}
\Omega_{h^0}=2g(T_\gamma)T^3_\gamma\frac{\sum_i m_i\Theta_i}{\rho_cg(T)},
\end{equation}
with
\begin{equation}\label{beta}
\Theta_i\equiv \frac{n_i}{T^3}=\frac{\eta \Gamma_i^2}{4 \pi^3 K m^3_{H^0_2}},
\end{equation}
where $T_\gamma = 2.4 \times 10^{-4}$ eV is the present photon temperature, $g(T_\gamma)=2$ is the photon degree of freedom, $g(T)=g_B+7g_F/8$ ($g_B$ and $g_F$ are the relativistic boson and fermion degrees of freedom, respectively), $\rho_c = 7.5 \times 10^{-47}h^2$ GeV$^4$ is the critical density of the Universe ($h \simeq 0.71$ is the Hubble constant in units of $100$ km s$^{-1}$ Mpc$^{-1}$), $\eta = 1.87$, $K^2=4 \pi^3 g(T)/45 m^2_{pl}$ and $m_{pl}=1.2\times 10^{19}$ GeV is the Planck mass. We will take $T=m_{H^0_2}$, since most of the contribution to each $\Theta_i$ comes from $T \leq m_{H^0_2} < T_{EW}$ \cite{mcdonald}.

The value of $g(m_{H^0_2})$ depends on the particle content of the particular 3-3-1 model under study, but typically $g(m_{H^0_2})\simeq 130$ for a three-family model \cite{FreTo}. For simplicity we take all the $m_i$ equal to a common value $m=5$ MeV, and $m_{H^0_2}=150$ GeV. Then, from Eqs.~(\ref{cosmic}) and (\ref{beta}) we get $\Omega_{h^0_2}\simeq 0.3$ for $\sum_i \Gamma_i^2 \simeq 1.88\times 10^{-12}$ GeV$^2$. Assuming that each $\Gamma_i$ contributes with approximately the same value to $\sum_i \Gamma_i^2$, we obtain $\Gamma_i \simeq 10^{-6}$ GeV, in broad agreement with the estimates in~\cite{mcdonald}, implying that such singlet scalars are very weakly coupled to the SM sector. Also in this case, as can be seen from Eq.~(\ref{acoples}), this value for $\sum_i \Gamma_i^2$ can be tuned with a variety of values of $u$ and $v$ and of the $\lambda$'s involved, a task that can be done more easily by taking, for example, $\lambda_7<0$ without spoiling the scalar mass spectrum. Consequently, the three light singlet scalars $h^0_2$, $H^0_3$ and $H^0_4$ do not overclose the Universe.

In conclusion, for 3-3-1 models without exotic electric charges we have considered the scalar sector and we have shown that for all these models this sector provides three light $SU(2)$ singlet neutral Higgs bosons, two CP-even and one CP-odd, each one with a mass in the range of a few MeV, and each one satisfying the properties of a candidate for thermally generated self-interacting dark matter in the Universe, all of them at the expense of tuning the values of combinations of several of the parameters in the scalar potential. These scalars have feeble interactions with ordinary matter, do not overpopulate the Universe and are stable without the need of an extra discrete symmetry.

The same problem was studied in Ref.~\cite{FreTo} for the PPF model, where the authors identify a light CP-odd scalar as a dark matter candidate at the price of introducing a new very small energy scale, different from $V$, $u$ and $v$, and of the order of $10^{-7}$ GeV, associated to the trilinear coupling $f$ in the scalar potential. A related analysis is done in~\cite{LoLa} for the so-called ``3-3-1 model with right-handed neutrinos'' (a model without exotic electric charges) when the discrete symmetry $\Phi_1 \rightarrow -\Phi_1$ is imposed, with the result that the model provides two scalar dark matter candidates, one CP-even and one CP-odd, also at the price of fine-tuning the parameters of the scalar potential. However, it is clear from Eq.~(\ref{vphi}) that this discrete symmetry forbids, among others, the trilinear coupling $f$, which plays a fundamental role in our analysis. So, what we have done is to generalize the study presented in Ref.~\cite{LoLa} by considering the most general scalar potential produced by the three Higgs scalar triplets of 3-3-1 models without exotic electric charges, without the introduction of an extra symmetry to account for the stability of the dark matter and that drastically reduces the scalar potential, and avoiding the introduction of a new mass scale. We also want to emphasize that our analysis is valid for all the 3-3-1 models without exotic electric charges, and not only for the so-called ``model with right-handed neutrinos".

The main conclusion of this study is that models based on the 3-3-1 gauge structure can provide good candidates for thermally generated self-interacting scalar dark matter but at the price of: introducing a new mass scale as done in~\cite{FreTo}, or introducing an ad-hoc discrete symmetry as done in~\cite{LoLa}, or by tuning some parameters of the scalar potential as shown in this letter.

\section*{ACKNOWLEDGMENTS}
Work partially supported by DIME at Universidad Nacional de Colombia-Sede Medell\'\i n, by CODI at Universidad de Antioquia and by I.C.R.A. (International Center for Relativistic Astrophysics) at University of Rome ``La Sapienza''..



\begin{thebibliography}{}

\bibitem[1]{nbody1}
B. Moore \textit{et al.}, Astrophys. J. {\bf 524}, L19 (1999); A.A. Klypin \textit{et al.}, Astrophys. J. {\bf 522}, 82 (1999).

\bibitem[2]{nbody2}
S. Ghigna \textit{et al.}, Astrophys. J. {\bf 544}, 616 (2000); J.F. Navarro, C.S. Frenk and S.D.M. White, \textit{ibid} {\bf 462}, 563 (1996); B. Moore \textit{et al.}, M.N.R.A.S. \textbf{310}, 1147 (1999).

\bibitem[3]{Spergel}
D.N. Spergel and P.J. Steinhardt, Phys. Rev. Lett. \textbf{84}, 3760 (2000).

\bibitem[4]{Yoshida}
N. Yoshida \textit{et al.}, Astrophys. J. \textbf{544}, L87 (2000); D.A. Buote \textit{et al.}, Astrophys. J. \textbf{577}, 183 (2002).

\bibitem[5]{astroph}
E. D'Onghia and A. Burket, Astrophys. J. \textbf{586}, 12 (2003).

\bibitem[6]{mcdonald}
J. McDonald, Phys. Rev. Lett \textbf{88}, 091304 (2002).

\bibitem[7]{burgess}
J. McDonald, Phys. Rev. D \textbf{50}, 3637 (1994); C.P. Burgess, M. Pospelov and T. ter Veldhuis, Nucl. Phys. B \textbf{619}, 709 (2001); M.C. Bento, O. Bertolami, R. Rosenfeld and L. Teodoro, Phys. Rev. D \textbf{62}, 041302 (2000); V. Silveira and A. Zee, Phys. Lett. B \textbf{161}, 136 (1985).

\bibitem[8]{FreTo}
D. Fregolente and M.D. Tonasse, Phys. Lett. B \textbf{555}, 7 (2003).

\bibitem[9]{LoLa}
H.N. Long and N.Q. Lan, Europhys. Lett. \textbf{64}, 571 (2003).

\bibitem[10]{pf}
F. Pisano and V. Pleitez, Phys. Rev. D \textbf{46}, 410 (1992);  P.H. Frampton, Phys. Rev. Lett. \textbf{69}, 2887 (1992).

\bibitem[11]{foot}
J.C. Montero, F. Pisano and V. Pleitez, Phys. Rev. D \textbf{47}, 2918 (1993); R. Foot, H.N. Long and T.A. Tran, Phys. Rev. D \textbf{50}, R34 (1994); H.N. Long, Phys. Rev. D \textbf{53}, 437 (1996).

\bibitem[12]{pospelov}
C. Bohem, D. Hooper, J. Silk and M. Casse, Phys. Rev. Lett. \textbf{92}, 101301 (2004): C. Picciotto and M. Pospelov, Phys. Lett. B \textbf{605}, 15 (2005).

\bibitem[13]{wandelt}
B.D. Wandelt \textit{et al.}, arXiv: astro-ph/0006344; W. de Boer, C. Sander, V.  Zhukov, A.V. Gladyshev and D.I. Kazakov, arXiv: astro-ph/0508617.

\bibitem[14]{valle}
M. Singer, J.W.F. Valle and J. Schechter, Phys. Rev. D \textbf{22}, 738 (1980).
\bibitem[15]{pfs}
W.A. Ponce, J.B. Fl\'orez, and L.A. S\'anchez, Int. J. Mod. Phys. A \textbf{17}, 643 (2002); W.A. Ponce, Y. Giraldo and L.A. S\'anchez, Phys. Rev. D \textbf{67}, 075001 (2003).

\bibitem[16]{spm}
L.A. S\'anchez, W.A. Ponce, and R. Mart\'\i nez, Phys. Rev. D \textbf{64}, 075013 (2001).

\bibitem[17]{mps}
R. Mart\'\i nez, W.A. Ponce and L.A. S\'anchez, Phys. Rev. D \textbf{65},
055013 (2002).

\bibitem[18]{ozer}
M. \"Ozer, Phys. Rev. D \textbf{54}, 1143 (1996).

\bibitem[19]{sher}
 D.L. Anderson and M. Sher, arXiv: hep-ph/0509200.

\bibitem[20]{long}
H.N. Long, Mod. Phys. Lett. A \textbf{13}, 1865 (1998).

\bibitem[21]{dmo}
R.A. Diaz, R. Mart\'\i nez and F. Ochoa, Phys. Rev. D \textbf{69}, 095009 (2004).

\end{thebibliography}
\end{document}